\documentclass[aps,prb,onecolumn,showpacs,superscriptaddress,groupedaddress,longbibliography]{revtex4-1}
\usepackage[dvips]{graphics,graphicx}
\usepackage{subfig}
\usepackage{bbold}
\usepackage{amsmath}
\usepackage{multirow}
\usepackage{amsfonts}
\usepackage{bm}
\usepackage{mathrsfs}
\usepackage{epstopdf}
\usepackage{hyperref}
\newcommand{\scs}{\scriptscriptstyle}

\begin{document}

\title{Refractive index sensing with optical bound states
in the continuum}

\author{Dmitrii N. Maksimov$^{1,2}$}
\author{Valeriy S. Gerasimov$^{1,3}$}
\author{Silvia Romano$^{4}$}
\author{Sergey P. Polyutov$^{1}$}

\affiliation{$^1$Siberian Federal University, 660041, Krasnoyarsk, Russia\\
$^2$Kirensky Institute of Physics, Federal Research Center KSC SB RAS, 660036, Krasnoyarsk, Russia\\
$^3$Institute of Computational Modeling SB RAS, Krasnoyarsk, 660036, Russia\\
$^4$Institute of Applied Sciences and Intelligent Systems, National Research Council, Naples, 80131, Italy}

\date{\today}
\begin{abstract}
We consider refractive index sensing with optical bounds states in the
continuum (BICs) in dielectric gratings. Applying a perturbative approach
we derived the differential sensitivity and the figure of merit of a sensor operating
in the spectral vicinity of a BIC. Optimisation design approach for engineering an
effective sensor is proposed. An analytic formula for the maximal sensitivity with an optical BIC is derived.
 The results are supplied with straightforward numerical simulations.
\end{abstract}
 \maketitle

\section{Introduction}
Bound states in the continuum (BICs) have revolutionized nanophotonics offering the opportunity to realize a new class of high throughput sensing devices and improving the control of the interaction between light and matter at the nanometer scale \cite{Molina2012, Romano18, BS1, Porter2005, Penzo2017, Zhen2014, Mocella2015, koshelev2019meta, Koshelev19, Zito2019}.
Strictly speaking, a BIC can be considered as a resonant mode with an infinite quality factor ($Q$-factor) in an open-cavity. This special mode has a frequency in the radiation continuum but does not lose energy because of symmetry mismatch  with radiative waves. The BICs are hosted by a leaky band of high{-}quality resonances with $Q$-factor diverging in the $\Gamma$-point \cite{Yuan17}. The leaky band with diverging $Q$-factor, in turn, induces a collapsing Fano feature in the transmittance spectrum \cite{Shipman,SBR,Blanchard16,Bulgakov18b}. Experimentally, this results in sharp spectral features in the transmission spectrum. Generally, the position of these extremely narrow Fano resonances is sensitive to the refractive index of the surrounding medium allowing to engineer optical sensors with a good sensitivity, \textit{S} and an excellent figure of merit (\textit{FOM}). Among the different geometries of structures supporting BICs, all-dielectric resonant planar structures, including photonic crystal slabs, periodic arrays and metasurfaces, have been taken as an alternative to conventional plasmonic sensors due to the possibility to realize high-performance sensing systems in loss-free media with a remarkable sensitivity to a small volume of the analyte \cite{Liu17,Romano18b, Romano18a, Yesilkoy19, Vyas20}. While \textit{S} is affected by the spatial overlap between the nonradiating evanescent field and the surrounding cladding, \textit{FOM} is proportional to the $Q$-factor and ultimately represents the sensor capability to follow tiny changes in the environment refractive index \cite{Romano19,Beheiry10}.  \\
In this paper we propose the derivation of the analytical expressions for \textit{S} and \textit{FOM} for sensors based on dielectric grating (DGs), one of the major set-ups for studying optical BICs both theoretically \cite{Marinica08, Monticone17, Bulgakov18, Bulgakov18a, Lee19, Bykov19, Gao19} and experimentally  \cite{Sadrieva17, Hemmati19}. By using a perturbative approach we find the differential sensitivity and the figure of merit of a DG-sensor operating in the spectral vicinity of a BIC. The reported analytic results could pave a way for an optimisation design approach for engineering an effective sensor based on BICs and provide a tool for a better understanding of the physics underlying this sensing mechanism.

\begin{figure*}[t]
\includegraphics[width=1\textwidth,trim={0cm 1.5cm 0cm 1cm},clip]{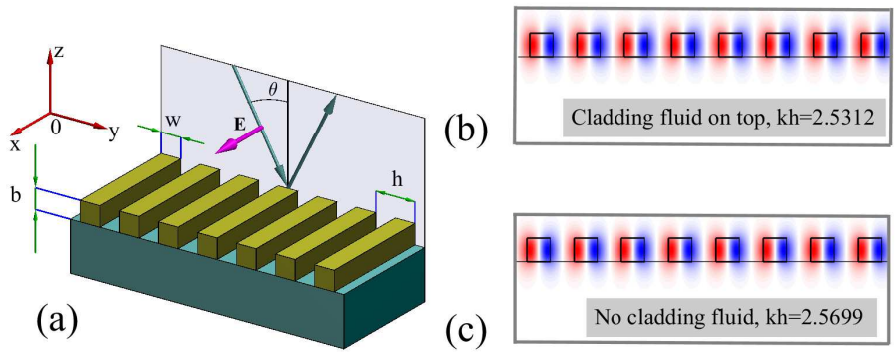} \caption{(a)
The dielectric grating assembled of rectangular bars on glass substrate. The plane of incidence $y0z$ is shaded grey.
The magenta arrow shows the electric vector of the incident wave. The geometric parameters are $w=0.5h$, $b=0.5h$.
 (b, c) BIC eigenmode profiles for Si bars as $E_y$ in the $y0z$-plane with ($n_c^{\scs (0)}=1.333$)
 and without ($n_c^{\scs (0)}=1$) cladding, correspondingly.} \label{fig1}
\end{figure*}

\section{Bound states in the continuum}
Here we consider the system shown in Fig.~\ref{fig1}~(a). It is a dielectric grating (DG) assembled of rectangular bars. The bars are periodically placed on the glass substrate with period $h$.  We assume that the incident field is TE polarized, i.e.  the electric vector is aligned with the bars as shown in Fig.~\ref{fig1}~(a). The propagation of the TE-modes is controlled by the Helmholtz equation for the $x$ component of the electric field:
\begin{equation}\label{Helmholtz}
\left(\frac{\partial^2}{\partial y^2}+ \frac{\partial^2}{\partial z^2}\right)E_x+k^2\epsilon E_x=0,
\end{equation}
where $k$ is the vacuum wavenumber, and $\epsilon$ is the dielectric function. In what follows we assume that the bars can be made of different dielectric materials such as
polycrystalline silicon, Si; and titanium dioxide, ${\rm Ti0_2}$.  The refractive index (RI) of the substrate $n_0=1.5$.

The RI sensor operating principle is detection of the shift in the wavelength of an optical resonance, $\lambda_{\rm res}$ in response to the change of
the RI of the cladding fluid, $n_c$ on top of the DG. As the reference value of the
cladding RI we take that of water $n_c^{\scs (0)}=1.333$. Our finite-difference time-domain (FDTD) simulations demonstrate that for the set of geometric parameters specified in the caption to Fig.~\ref{fig1} a Si DG supports a BIC with $kh=2.5312$. The BIC is observed as a specific
point of the leaky band dispersion with infinite life-time of the resonant mode in the $\Gamma$-point (see e.g. \cite{Bulgakov18b}).
The field
profile of the BIC in the Si DG with cladding fluid is shown in Fig.~\ref{fig1}~(b). The BIC clearly falls into the symmetry protected type as it is symmetrically mismatched to the zeroth order diffraction channel \cite{Bulgakov18b}. It is remarkable that
the BIC is robust with respect to variations of $n_c$ as it can be seen from Fig.~\ref{fig1}~(c), where we demonstrate the field profile of the BIC at $kh=2.5699$ with no cladding fluid. Notice that the mode profiles are indistinguishable to the naked eye. The simulations were also run for a DG made of ${\rm Ti_20}$. The obtained BIC field profiles
are almost identical to those in Fig.~\ref{fig1}~(b) and  Fig.~\ref{fig1}~(c), and, hence, are not shown in the paper.
The numerical data for both DG materials are collected in Table~\ref{table1}.
\begin{table}[h]
 \caption{BICs in dielectric gratings. The values of RI are taken at 1$\mu$m. The BIC
 wavelenghts, $\lambda_{\rm \scs BIC}$ are measured in the units of $h$.}
 \begin{center}
\begin{tabular}{|l|ccccc|}
\hline
{Grating~material} & Si  &Si & ${\rm Ti0_2}$ & ${\rm Ti0_2}$ &
\\ \hline
Grating~{RI}  & 3.575 & 3.575 & 2.485 & 2.485 &
\\ \hline
Cladding~fluid  & Yes  & No & Yes & No &
\\ \hline
$hk$ & 2.5312  & 2.5699 & 3.4479  & 3.5637  &
  \\ \hline

$\lambda_{\rm \scs BIC}$, (p.d.u.) & 2.4823h  &  2.4449h & 1.8223h  & 1.7631h  &
  \\ \hline
\end{tabular}
\end{center}
\label{table1}
\end{table}

\section{Refractive index sensing}
Although the BIC proper is
totally decoupled from the outgoing waves, it is spectrally surrounded by a leaky band with with a diverging $Q$-factor. If the DG is illuminated from the far-zone, this leaky band induces a collapsing Fano feature in the transmittance spectrum. Below we investigate into application of
the collapsing Fano resonance for RI sensing. Two quantities are of major importance for engineering an effective sensor: the
differential sensitivity defined as
\begin{equation}\label{s}
    S=\left.\frac{d\lambda_{{\rm res}}}{d n_c}\right|_{n_c=n_c^{(0)}},
\end{equation}
and the figure of merit (FOM) given by
\begin{equation}\label{FOM}
 {\rm FOM}=\frac{S}{W},
\end{equation}
where $W$ is the width of the operating resonance in terms of wavelength.

Here we aim at deriving analytic expressions for both quantities of interest
using a perturbative approach. The perturbative approach utilizes two small parameters: the first is the increment
of the dielectric function $\Delta \epsilon_c$ due to the change of the RI of the cladding, and the second is the angle of incidence $\theta$ defined in
Fig.~\ref{fig1}~(a). First of all let us write the solution
of Eq.~\eqref{Helmholtz} as
\begin{equation}
E_x(y,z)=\psi(y,z) e^{i\beta y},
\end{equation}
where $\beta=k\sin(\theta)$ is the propagation constant along the $y$-axis.
After substituting the above into Eq.~\eqref{Helmholtz} and taking into
account $\Delta \epsilon_c$ one finds
\begin{equation}
    \left(\frac{\partial^2}{\partial y^2}+ \frac{\partial^2}{\partial z^2}\right)\psi+ k^2(\mathscr{L}_c+\mathscr{L}_{\theta})\psi+  k^2\epsilon \psi=0,
\end{equation}
where the perturbation operators are given by
\begin{align}
    & \mathscr{L}_c=\Delta \epsilon_c, \\
    & \mathscr{L}_{\theta}=2i\frac{\sin(\theta)}{k}\frac{d}{dx}+
    \sin^2(\theta)\frac{d}{dx^2}.
\end{align}
Notice that $\mathscr{L}_c$ and $\mathscr{L}_{\theta}$ are independently
parametrized with respect to $n_c$ and $\theta$. It means that
the contributions of the two operators into the perturbed eigenfrequency
are additive in the first perturbation order.

First let us see the effect of $\mathscr{L}_c$. In general, the application of perturbative approaches to resonant states in not trivial, since the
eigenfields diverge in the far-zone \cite{zel1961theory, lai1990time, Lalanne18}. One of the possible solutions is the use of the Wigner-Brillouin (WB) perturbation theory \cite{Muljarov10, Lobanov18}. The application of the WB
approach requires a very specific normalization condition for the resonant
states involving both "volume" convolution with dielectric function, and "flux" term as a surface integral over the outer interface of the elementary cell. It can be argued, however, that the BIC proper is
a specific case when the flux term can be dropped off due to the eigenfield
vanishing in the far-zone \cite{pankin2020fano}. Thus, for a single BIC mode the WB  perturbation theory yields vacuum wavenumber
\begin{equation}\label{solutionWB}
    k=\frac{2k^{\scs (0)}}{2+I_c\Delta \epsilon_c},
\end{equation}
where $k^{\scs (0)}$ is the vacuum wavenumber of the BIC with
the reference value of the cladding RI and
\begin{equation}\label{Ia}
I_c=\int_{{\cal S}_c} \left(E_x^{\scs \rm (BIC)}\right)^2 dydz
\end{equation}
with ${\cal S}_c$ as the area of the elementary cell occupied by the cladding fluid, and $E_x^{\scs \rm (BIC)}$ as the BIC eigenfield satisfying the following normalization condition
\begin{equation}\label{norm}
    1=\int_{{\cal S}} \left(E_x^{\scs \rm (BIC)}\right)^2 \epsilon(y,z) dydz,
\end{equation}
where ${\cal S}$ is the total area of the elementary cell. Notice that although
the integration domains are infinite both integrals converge due to the eigenfield vanishing in the far-zone.

The normalization condition~\eqref{norm} naturally invites applying the standard  Rayleigh-Schr\"odinger (RS) perturbative approach following \cite{Mortensen07}. Remarkably, unlike the
WB method the first order RS approach does not involve any other states
rather than the unperturbed state under consideration \cite{Landau58a}. The first order RS solution reads
\begin{equation}\label{solutionRS}
    k=k^{\scs (0)}\left(1-\frac{1}{2} I_c\Delta \epsilon_c\right)+{\cal O}(\Delta\epsilon_c^2).
\end{equation}
We stress again that
although the solution \eqref{solutionWB} does not require smallness of $\Delta \epsilon_c$, it is still approximate since it neglects contribution
of all unperturbed modes other than the BIC. In contrast the first order non-degenerate
RS solution \eqref{solutionRS} only involves a single mode but
requires smallness of $\Delta \epsilon$. Notice though, that Eq. \eqref{solutionWB} and Eq. \eqref{solutionRS} coincide up to ${\cal O}(\Delta\epsilon_c^2)$. Thus, the Wigner-Brillouin approach truncated
to a single eigenstates produces an exact result up to ${\cal O}(\Delta\epsilon_c^2)$.

Unfortunately, unlike $\mathscr{L}_c$ the effect of $\mathscr{L}_{\theta}$
can not be described by the single state WB approach. This is because
the perturbed solution is a radiating state and, thus, even in the first
order approximation can not be written through the non-radiating unperturbed BIC. Physically, the application of $\mathscr{L}_{\theta}$ generates the dispersion of the leaky band hosting the BIC in the $\Gamma$-point. Since the band is symmetric with respect to  $\beta\rightarrow-\beta$ the dispersion about the $\Gamma$-point reads
\begin{equation}\label{dispersion}
    k=k^{\scs (0)}(1+\alpha\theta^2)+ {\cal O}(\theta^4).
\end{equation}
The complex-valued parameter $\alpha$ can be found by fitting to numerical
or experimental data. As it was already mentioned the corrections due to
$\mathscr{L}_c$ and $\mathscr{L}_{\theta}$ are additive in the leading order
because both $\theta$ and $\Delta n_c$ are regarded as small quantities.
Thus, using $\lambda=2\pi/k$, Eq.~\eqref{solutionWB}, and Eq.~\eqref{dispersion} we obtain the resonant wavelength, $\lambda_{\rm res}$ as
\begin{equation}\label{lambda}
\lambda_{\rm res}=\lambda_{\scs \rm BIC}^{\scs (0)}\left(1+\frac{1}{2}I_c\Delta\epsilon_c-\alpha\theta^2\right)
\end{equation}
Notice that Eq.~\eqref{lambda} explicitly predicts blue shift of the resonant
wavelength with positive $\Delta\epsilon_c$.
Applying the definition of the sensitivity, Eq.~\eqref{s} one finds
\begin{equation}\label{sensitivity}
    S=\lambda_{\scs \rm BIC}^{\scs (0)}n_c^{\scs (0)}I_c.
\end{equation}
The above expression for sensitivity exactly coincides with that previously
derived by Mortensen, Xiao, and Pedersen \cite{Mortensen07} for guided modes in a bulk photonic crystal.

Remarkably, Eq.~\eqref{sensitivity} can also be derived with
a straightforward application of the Hellmann-Feynman (HF)
 theorem \cite{Politzer18}. Notice, though, that the HF theorem strictly requires the problem to be described
by Hermitian operators. In general the radiating boundary conditions
break the hermiticity. However, in the case of the BIC proper the eigenfield is localized about the DG and, thus, is not affected by the boundary condition in the far-zone.

To derive the FOM we recall that the width $W$ of a high-$Q$ resonance can be written as
\begin{equation}\label{W}
    W=\frac{\lambda^{\scs (0)}_{\scs \rm BIC}}{Q},
\end{equation}
where $Q$ is the quality factor
\begin{equation}\label{Q}
    Q=\frac{k^{\scs (0)}}{2\Im\{a\}}.
\end{equation}
Hence,
for the FOM we have
\begin{equation}\label{FOM_final}
    {\rm FOM}=n_c^{\scs (0)}I_cQ\propto \theta^{-2}.
\end{equation}

\begin{figure*}[t]
\centering\includegraphics[width=1.\textwidth,trim={0cm 0cm 0cm 0cm},clip]{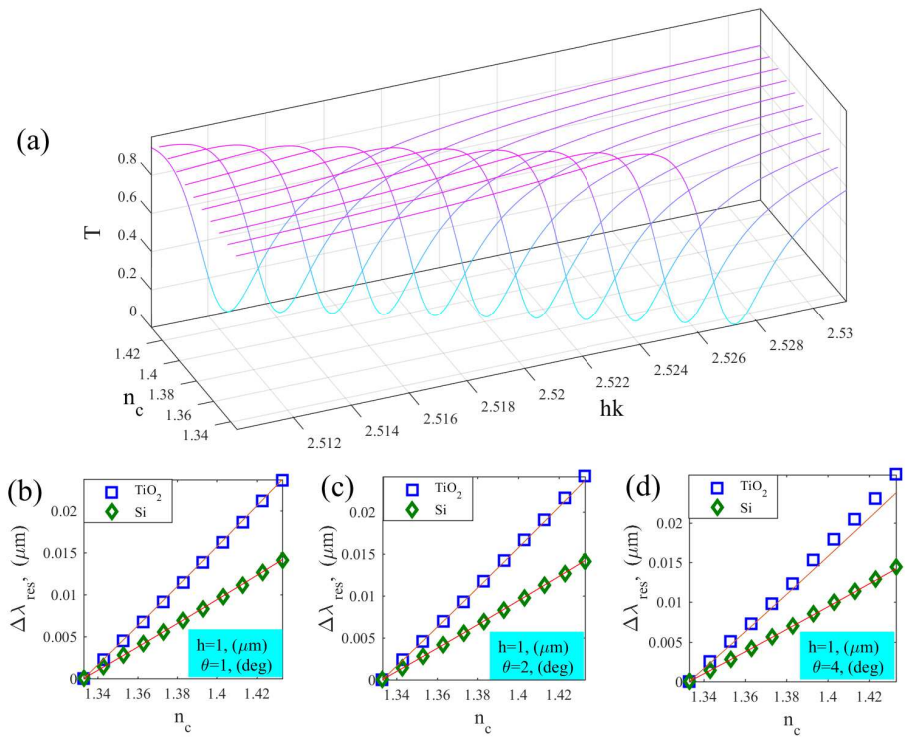}
\caption{Blue shift of the resonant frequency. (a) Fano feature in transmittance, $T$ against the incident wave number, $kh$ and the refractive index of the cladding fluid, $n_c$; the DG is made of $Si$ and illuminated at the angle of incidence $\theta=4$, (deg). (b-d) Shift of the
resonant wavelength, $\Delta\lambda_{\rm res}$ as a function of $n_c$ at
three different angles of incidence; DG material and $\theta$ are specified in the subplots.}
\label{fig2}
\end{figure*}
\begin{table}[h]
 \caption{Numerical and theoretical values of sensitivity. RIU stands for refractive index unit.}
 \begin{center}
\begin{tabular}{|l|cccccc|}
\hline
{Grating~material} & Si  & Si & Si & ${\rm Ti0_2}$ & ${\rm Ti0_2}$ & ${\rm Ti0_2}$
\\ \hline
$\theta$, (deg) & 1 & 2 & 4 & 1 & 2 & 4
\\ \hline
$S_{\rm theor}$, (nm/RIU)  & 137.2 & 137.2  & 137.2  & 229.7  & 229.7 & 229.7
\\ \hline
 $S_{\rm num}$, (nm/RIU)  & 137.4  & 137.5  & 137.8  & 222.4  & 233.7  & 252.3 \\ \hline
$\delta$                      & 0.037\% & 0.052\% & 0.114\% & 0.812\% & 0.429\% &  2.343\%
  \\ \hline
\end{tabular}
\end{center}
\label{table2}
\end{table}

Finally, let us discuss the upper limit of the differential sensitivity. By comparing the integral~\eqref{Ia} against Eq.~\eqref{norm} one writes
\begin{equation}\label{integral}
    I_c=\frac{1}{\epsilon_c}
    \left[ 1-\int_{{\cal S}_{\scs \rm DG}}\epsilon(y,z) \left(E_x^{\scs \rm (BIC)}\right)^2dydz \right],
\end{equation}
where ${\cal S}_{\scs \rm DG}$ is the area of the elementary cell occupied by the DG.
The quantity $I_c$ reaches its maximal value, when
the integral in Eq.~\eqref{integral} equals to zero.
Thus, the maximal differential sensitivity in the spectral vicinity of an isolated BIC
is given by
\begin{equation}\label{limit}
    S_{\rm max}=\frac{\lambda_{\scs \rm BIC}}{n_c}.
\end{equation}

\section{Numerical results}
To verify the above findings we run FDTD simulations for computing the transmittance, $T$ as a function of both
$k$ and $\theta$. In Fig.~\ref{fig2}~(a) we demonstrate the blue shift of the Fano feature with the increase of the cladding RI.
In Figs.~\ref{fig2}~(c-d) we plotted the shift of the resonant wavelength $\Delta\lambda_{\scs \rm res}=\lambda^{\scs (0)}_{\scs \rm res}-\lambda_{\scs \rm res}$ in comparison against
Eq.~\eqref{lambda} at three different values of $\theta$. On can see from Figs.~\ref{fig2}~(c-d) that the data match to a good accuracy with deviation only becoming visible at a larger angle of incidence, $\theta=4$, (deg), when the higher perturbation orders in $\theta$ come into play.

In Table~\ref{table2} we present the numerical values of differential sensitivity
obtained from Figs.~\ref{fig2}~(c-d) in comparison against the analytic result Eq.~\eqref{sensitivity}. To assess the accuracy of Eq.~\eqref{sensitivity} we calculated
the relative error
\begin{equation}\label{error}
\delta=\frac{|S_{\rm theor}-S_{\rm num}|}{2(S_{\rm theor}+S_{\rm num})}\cdot 100 \%,
\end{equation}
where $S_{\rm theor}$ stands for the theoretical value of the sensitivity found from
Eq.~\eqref{sensitivity}, while $S_{\rm num}$ is used for numerical data.
Again one can see a good coincidence with $\delta>1\%$ only for $\theta=4$, (deg)
with a $\rm TiO_2$ DG. Otherwise the sensitivity remains constant about the normal incidence as predicted by Eq.~\eqref{sensitivity}.

\begin{figure*}[h]
\centering\includegraphics[width=1.\textwidth,trim={0cm 0cm 0cm 0cm},clip]{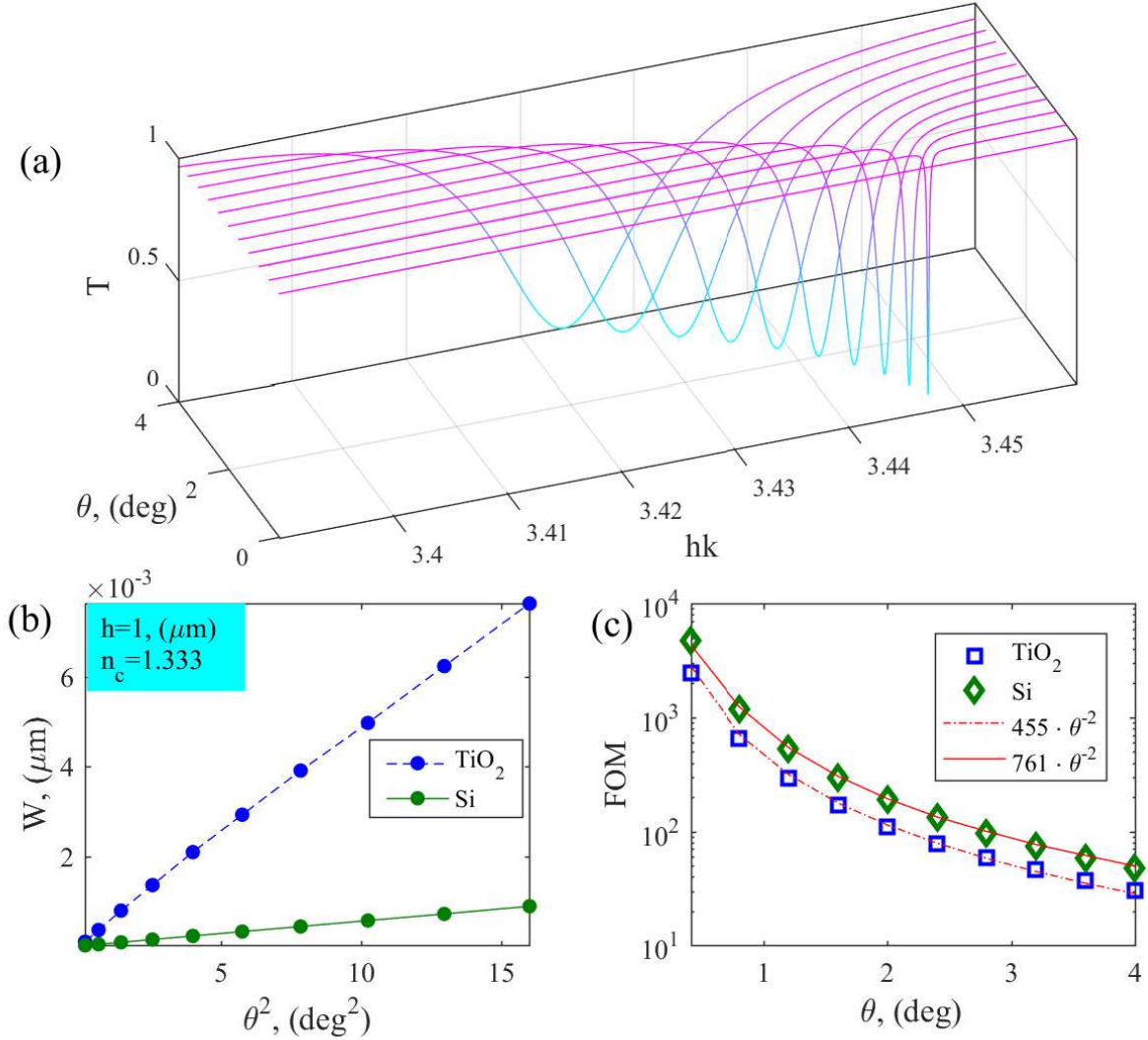}
\caption{Fano resonance and FOM. (a) Collapsing Fano feature in transmittance, $T$ on approach to
the normal incidence in $\rm TiO_2$ DG; $n_c=1.333$. (b) Resonant width in terms of wavelength against $\theta^2$. (c) FOM as a function of the angle of incidence. Thin
lines show fitting to ${\rm FOM} \propto \theta^{-2}$.}
\label{fig3}
\end{figure*}
In Fig.~\ref{fig3}~(a) we show the Fano feature in transmittance collapsing
on approach to normal incidence. The width of the resonances is extracted from
the data in Fig.~\ref{fig3}~(a) and plotted against $\theta^2$ in Fig.~\ref{fig3}~(b).
One can see from Fig.~\ref{fig3}~(b) that the numerical data comply with
Eq.~\eqref{W} and Eq.~\eqref{Q} which predict $W\propto \theta^2$. Finally,
in Fig.~\ref{fig3}~(c) we plot the FOM against the angle of incidence together
with fitting curves ${\rm FOM} \propto \theta^{-2}$. One can see that the numerical results are in accordance to Eq.~\eqref{FOM_final}.

\section{Summary and conclusions}
The equations derived provide a cue to designing an efficient sensor with an optical BIC. The expression for sensitivity obtained in this paper is identical to that derived by Mortensen, Xiao, and Pedersen \cite{Mortensen07} for bulk photonic crystal underlying the unique properties of BICs among the leaky modes supported by the grating.
According to Eq~\eqref{sensitivity}, and Eq.~\eqref{integral}  the obvious approach
for maximising the sensitivity is either manufacturing the DG of a dielectric with small $\epsilon$ or choosing a BIC with the eigenfield predominantly occupying the free space
above the DG. Both design paradigms can be implemented by
solving the eignevalue problem with FDTD method with variation of the geometric parameters.

On the other hand, the FOM can be optimized by operating at near normal incidence. At this point some comments are due in regard to FOM, Eq.~\eqref{FOM_final} diverging with
the decrease of the angle of incidence. Experimentally, a zero linewidth can not obtained in a realistic set-up due
to three major factors:

(i) material losses due to absorption in the dielectric \cite{Sadrieva19, Hu20};

(ii) fabrication inaccuracies \cite{Ni17}; and

(iii) finiteness of the structure \cite{Bulgakov17oe, Sadrieva19}, as the BIC proper can not be supported by systems finite in all spatial dimensions \cite{Silveirinha14}.

The effect of all three factors is saturation of FOM on
approach to normal incidence. Which of the three will
dominate in the staturation is determined by a specific design of the DG.

Another important aspect is the finite waist of the Gaussian beam illuminating the DG. Equations~\eqref{sensitivity}, and~\eqref{FOM_final} can only be strictly applied to a plane wave, whereas a Gaussian beam is always
a continuum of plane waves propagating to slightly different directions. Miniaturization of the sensor would
imply a tightly focused beam with a broad momentum distribution in the Fourier plane. This would naturally compromise the FOM, as the scattering channels with larger angles of incidence become more populated.

Finally, let us discuss whether the limit, Eq.~\eqref{limit} can be overcome. First of all, Eq.~\eqref{limit} is derived under assumption of smallness of the perturbation operators. Therefore, the sensitivities
larger than Eq.~\eqref{limit} are not strictly forbidden.
For example, a larger sensitivity has been recently reported with surface plasmon-polaritons \cite{Bikbaev20}. The drawback, however, is the drop of the quality factor due to strong coupling of the eigenmode to the outgoing waves. Thus, the increase of sensitivity with departure from the BIC proper can be only obtained at the cost of the sensor FOM. More promising is optimizing the sensitivity while keeping the high FOM intact. This can be done by breaking the second condition
in derivation of Eq.~\eqref{limit}, namely, the single mode approach. One possible route is looking for a BIC in the exceptional points \cite{El-Ganainy19, Oezdemir19, Krasnok19}, which unavoidably emerge as a result of an intricate interplay between two eigenmodes. The other possible route is application of BIC existing just below the first Wood's anomaly in which case the contribution of the
evanescent channels corresponding to the first diffraction
order becomes extremely sensitive to the system's parameters \cite{Romano19}. The destruction of a BIC in crossing the Wood's
anomaly inherently implies application of a multimodal perturbative approach since the contribution of states with finite live-time is required. We speculate that operating in the vicinity of the Wood's anomaly may result in a significant
gain in sensitivity while keeping diverging the FOM and quality factor.

\section*{Acknowledgments.} The research was supported by the Ministry of  Science  and  Higher  Education  of  Russian  Federation, project no.  FSRZ-2020-008. DNM is grateful to Andrey A. Bogdanov for useful discussion.

\section*{Disclosures.} The authors declare no conflicts of interest.

\bibliography{BSC_light_trapping}


\end{document}